%% file: main27_aps.tex
\documentclass[aps,prd,twocolumn,showpacs,preprintnumbers,amsmath,amssymb,superscriptaddress]{revtex4}

\usepackage{graphicx}
\usepackage{times}
\usepackage{color}

\input{./Texs/shorthand}


\input{./Texs/abb_info}

\graphicspath{{./Fig/}}

\newcommand{\re}{{\rm Re}}
\renewcommand{\Re}{{\rm Re}}

\renewcommand{\Im}{{\rm Im}}
\renewcommand{\mod}{{\rm mod}}

\newcommand{\comment}[1]{{\color{black} #1}}
\newcommand{\commentb}[1]{{\color{black} #1}}

\begin{document}

\title{
Lee-Yang zero distribution of high temperature QCD and Roberge-Weiss phase transition 
}

\author{Keitaro Nagata}
\email{knagata@post.kek.jp}
\affiliation{\affKEK}
\author{Kouji Kashiwa}
\email{kouji.kashiwa@yukawa.kyoto-u.ac.jp}
\affiliation{\affYITP}
\author{Atsushi Nakamura}
\email{nakamura@riise.hiroshima-u.ac.jp}
\affiliation{\affHURIISE}
\author{Shinsuke M. Nishigaki}
\email{mochizuki@riko.shimane-u.ac.jp}
\affiliation{\affSNU}

\date{\today}

\begin{abstract}
Canonical partition functions and Lee-Yang zeros of QCD at finite density and high temperature are studied.
\comment{Recent lattice simulations
have confirmed that the free energy of QCD is 
a quartic function of quark chemical potential at temperature 
slightly above pseudo-critical temperature $T_c$,
as in the case with a gas of free massless fermions.
}
We present analytic derivation of the canonical partition functions and Lee-Yang zeros  
for \comment{this type of} free energy using
the saddle point approximation.
We also perform lattice QCD simulation in a canonical approach
using the fugacity expansion of the fermion determinant, and carefully examine its reliability.
By comparing the analytic and numerical results, we conclude that the canonical partition functions 
follow the Gaussian distribution of the baryon number, and the accumulation of Lee-Yang zeros 
of these canonical partition functions exhibit the first-order Roberge-Weiss phase transition. 
We discuss the validity and applicable range of the result and its implications 
both for theoretical and experimental studies.
\end{abstract}
\maketitle

\section{Introduction}
Quantum chromodynamics (QCD) undergoes a phase transition from the hadronic phase to 
the quark gluon plasma (QGP) phase at high temperature.
Recently, a beam energy scan (BES) program at the Relativistic Heavy Ion Collider
has reported valuable data for the long-standing issue of identifying the phase boundary 
in the QCD phase diagram by using heavy ion collisions with different collision energies, centralities, etc.
\cite{Aggarwal:2010wy,Adamczyk:2013dal,Adamczyk:2014fia}.
There the probability distribution of conserved charges has been constructed by measuring them for each collision. 
Extensive efforts have been invested to understand event-by-event fluctuations of those charges, as
they are expected to be useful observables for locating the critical end point~\cite{Stephanov:1998dy,Hatta:2003wn,Stephanov:2008qz,Asakawa:2009aj,Stephanov:2011pb}.

The setup in BES experiments, where a part of fireballs 
is accessible for measurements, resembles a grand canonical ensemble
in statistical mechanics~\cite{BraunMunzinger:2011dn,Morita:2012kt,Garg:2013ata}, 
and this parallelism enables us to study the probability distribution
of the net baryon number $n_{\rm B}$ theoretically\comment{:}
Consider a grand canonical ensemble for
\comment{a single particle species}.
The grand canonical partition function $Z(\mu)$ is expanded in terms of the number of
particles $n$ as  $Z(\mu) = \sum_n Z_n e^{n \mu/T}$.
Here $Z_n$ is a canonical partition function, which depends on temperature $T$
but not on the chemical potential $\mu$.
For given $\mu$ and $T$, $Z_n e^{n \mu/T}$ is proportional to the probability
of observing an $n$-particle state in the grand canonical system.
The BES experiments have so far measured the net proton multiplicity and reported \cite{Adamczyk:2013dal} that
it closely follows the Skellam distribution for several collision energies and centralities.
This observation is consistent with the hadron resonance gas (HRG) model,
in which the net baryon multiplicity is approximately given by a Skellam distribution~\cite{BraunMunzinger:2011dn}.

On the other hand, in lattice QCD simulations the canonical approach
has been proposed as a tool to circumvent the sign problem associated with a finite chemical potential
~\cite{Barbour:1988ax,Barbour:1991vs,Hasenfratz:1991ax,deForcrand:2006ec,Kratochvila:2005mk,Kratochvila:2004wz,Ejiri:2008xt,Li:2010qf,Danzer:2012vw,Nagata:2012tc,Nakamura:2013ska}.
In previous studies~\cite{Nagata:2012tc,Nakamura:2013ska},
two of the authors (K. N. and A. N.) found that
the canonical partition functions follow the Gaussian distribution
of the baryon number at high temperatures and that
the Lee-Yang zeros obtained from the canonical partition functions of the Gaussian type
exhibit a behavior consistent with a Roberge-Weiss (RW) phase transition~\cite{Roberge:1986mm}. 
This result has several implications\comment{:} 
The connection between the Gaussian behavior of the net baryon number distribution and the RW phase transition 
can be used as an experimental probe indicating the QGP phase, since the RW phase transition is a 
phenomenon specific to the QGP phase.
The result is also interesting in the context of the Lee-Yang zero analysis~\cite{Yang:1952be,Lee:1952ig}, 
since the distribution of Lee-Yang zeros is known for some limited cases, e.g.~\cite{Lee:1952ig,Biskup:2000prl1}.

Despite the aforementioned importance, the determination of Lee-Yang zeros 
in Monte Carlo simulations is \comment{a} difficult task.
Canonical partition functions suffer from a phase fluctuation configuration by 
configuration due to the sign problem. 
This problem may be reduced by using sophisticated approaches proposed in e.g.~\cite{deForcrand:2006ec,Li:2010qf}.
However, the phase fluctuation becomes more severe as the baryon number increases, 
and the truncation of the fugacity polynomial at a certain order is inevitable.  
Such methodological artifacts might sensitively affect the thermodynamic behavior of the Lee-Yang zeros
that are the roots of the truncated polynomial.

The purpose of the present work is to determine the canonical partition functions and 
Lee-Yang zeros in QCD at high temperatures and to reexamine their relationships.
To this end, we present an alternative analytical calculation and 
assess the reliability of the lattice results in light of the former.
Specifically, we first derive the canonical partition functions and Lee-Yang zeros at high temperature
by utilizing the fact that the free energy is then given as a simple quartic
function of the quark chemical potential.
The canonical partition function is defined as a Fourier integral 
of the grand canonical partition function with pure imaginary chemical potential.
At high temperature, this integral can be 
evaluated in a saddle point approximation, 
yielding the Gaussian function. 
Accordingly, the grand canonical partition function is expressed
as a Jacobi theta function.
Using the property of the zeros of the theta function, we show that the Lee-Yang zeros are located on 
the negative real axis on the complex plane of the baryon fugacity. 
Because of the RW periodicity, these zeros are aligned on three radial lines on the complex plane
of the quark fugacity. 
This elucidates the close connection between the Gaussian behavior of the canonical partition functions 
and the RW phase transition. 

In reexamining the results from the lattice QCD simulations,
some of which have been already presented in ~\cite{Nakamura:2013ska},
we newly address the issue of the convergence of the fugacity polynomial and Lee-Yang zeros,
and perform a bootstrap analysis for their distribution.
We find that the Lee-Yang zeros related to the RW phase transition are not sensitive 
to the truncated part of the fugacity polynomial. 
We also find an agreement between the lattice data and analytic calculation. 

This paper is organized as follows.
In the next section, we explain the canonical approach and Lee-Yang zero theorem.
In Sec. \ref{sec:jtheta}, we explain some features of QCD. 
Using those features, we derive the canonical partition functions and Lee-Yang zeros. 
In Sec. \ref{sec:simulation}, we compute the canonical partition functions 
and Lee-Yang zeros in lattice QCD simulation. 
We also discuss implications and reliability of the results. 
The final section is devoted to a summary.

\section{Canonical approach and Lee-Yang zeros}

In this section, we explain the canonical approach and 
the Lee-Yang zero theorem. 
The grand canonical partition function is defined by 
\begin{align}
Z(\mu)= {\rm tr}\, e^{ - (\hat{H}- \mu \hat{N})/T},
\end{align}
where $\hat{H}$ and $\hat{N}$ denote the Hamiltonian and the quark number operator in QCD, 
and $T$ and $\mu$ the temperature and the quark chemical potential.
We refer to $V$ as the spatial volume of the system. 
Using the eigenstates of the number operator, $Z(\mu)$ can be expanded 
in powers of fugacity $\xi=e^{\mu/T}$, 
\begin{align}
Z(\mu) = \lim_{N\to\infty} \sum_{n=-N}^N Z_n \xi^n. 
\label{Eq:2014Mar16eq0}
\end{align}
Here, $Z_n= {\rm tr}( e^{-\hat{H}/T}\,\delta_{\hat{N},n})$
is the canonical partition function at a fixed quark number $n$, 
which is the eigenvalue of $\hat{N}$. 
$Z_n$ is real and positive for any $n$, and satisfies $Z_n = Z_{-n}$. 
$N$ is the maximum number of quarks that can be supported on the system. 
The maximum number $N$ is finite on the lattice and diverges in the thermodynamic limit.
\commentb{$Z_n$ is related to the Helmholtz free energy density $f_{\rm H} (n)$ as 
$Z_n = \exp(-V f_{\rm H} (n) /T)$~\cite{Kratochvila:2005mk,Ejiri:2008xt,Alexandru:2010yb}, }
and converges in the thermodynamic limit.
By extending 
$\mu$ to pure imaginary values, 
$\mu=i \mu_I, \mu_I \in \mathbb{R}$, Eq.~(\ref{Eq:2014Mar16eq0}) 
is  regarded as a Fourier expansion of $Z(\mu)$ with the 
Fourier coefficients $Z_n$. 
The latter can be expressed as the Fourier transformation  
\begin{equation}
Z_n 
= \int d \theta Z (\theta) e^{i n \theta}
     = \int d \theta e^{ - V f(\theta)/T} e^{ i n \theta}, 
\label{Eq:2014Dec04eq1}
\end{equation}
where $\theta = \mu_I/T$, and we have used $f(\mu)=-(T/V) \ln Z(\mu)$ to 
obtain the right-hand side~\footnote{We use Eq.~(\ref{Eq:2014Dec04eq1}) for analytic calculation, 
while we a fugacity expansion formula for numerical simulations. See Sec. IV A}.
\commentb{Here $f(\mu)$ denotes the Gibbs free energy density}.
The domain of the Fourier integral is usually $0$ to $2\pi$. 
In QCD, we need to take into account the Roberge-Weiss periodicity to determine 
the domain of the integral, 
as we elaborate in the next section.

For real $\mu$ (i.e.~ real and positive $\xi$), 
$Z(\mu)$ can never have zeros
since its coefficient $Z_n$ is real and positive for any $n$.
However, $Z(\mu)$ can have zeros for complex $\mu$. 
Using the roots $\xi_i$ of $Z(\mu)$ in the complex $\xi$ plane, 
it is expressed in a factorized form\comment{~\cite{Yang:1952be,Lee:1952ig}}
\begin{align}
Z(\mu) = \lim_{N\to \infty} Z_{-N} \xi^{-N}  \prod_{i=1}^{2N}
\left(1-\frac{\xi}{\xi_i}\right).
\label{Eq:2014Mar16eq1}
\end{align}
The roots $\{\xi_i\}$ are referred to as Lee-Yang zeros.
Because of the symmetry $Z_n = Z_{-n}$,
any root $\xi_i$ inside the unit circle is accompanied by another root $1/\xi_i$ outside.

In spite of its general importance, it is practically 
difficult to obtain Lee-Yang zeros for different models.
Lee and Yang ~\cite{Lee:1952ig} showed that Lee-Yang zeros in Ising models are distributed only on
the unit circle on the complex plane of $e^h$; see Fig.~\ref{Fig:2014Sep21fig1}.
\comment{
To relate Lee-Yang zeros to thermodynamic singularities, it is useful to 
recall an electrostatic analogue 
proposed by the very founders
\cite{Lee:1952ig}
and later used in the context of QCD
in e.g.~\cite{Blythe:2003aaa,Stephanov:2006dn,Ejiri:2014oka}.
Considering the free energy as an analytic function on the complex $\xi$-plane, we
denote its real part by $\phi \equiv \Re\, f$, which is written as
\begin{align}
\phi(\xi)
 = - \frac{T}{V} \sum_{i=1}^{2N} \ln|\xi-\xi_i| - \frac{T}{V} \ln Z_{N} + \frac{NT}{V} \ln |\xi|. 
\end{align}
Here the third term comes from a multiplicative factor of the grand canonical partition function, 
and is irrelevant to Lee-Yang zeros. 
This also provides a constant contribution to the number density because $\ln \xi=\mu/T$, and 
is irrelevant to phase transitions. 
Therefore, we can safely ignore this term~\cite{Stephanov:2006dn,Blythe:2003aaa}. 
Taking the derivatives  of $\phi$ with respect to $\xi$, we obtain
\begin{align}
\nabla_\xi^2
\phi(\xi) = - 2\pi \frac{T}{V} \sum_{i=1}^{2N}
\delta^{(2)}
(\xi-\xi_i), 
\label{Eq:2015Jan04eq1}
\end{align}
where $
\nabla_\xi 
\equiv ( \partial/(\partial\,\Re\,\xi), \partial/(\partial\,\Im\,\xi))$. 
We have used 
\begin{align}
\nabla^2
 \ln |z| = \left( \frac{\partial^2 }{\partial x^2} + 
\frac{\partial^2}{\partial y^2} \right) \ln | x+ iy| = 2 \pi \delta(x) \delta(y). 
\end{align} 
Equation~(\ref{Eq:2015Jan04eq1}) 
is just the Poisson's equation
for a two-dimensional 
electrostatic potential problem;  the
real part of the free energy $\phi$  is interpreted as the electrostatic potential, 
its derivative $\nabla_\xi \phi$
as the electric field, and Lee-Yang zeros as the location of 
charges. 
In an electrostatic problem
of charges that 
accumulate
e.g.~on a circle, 
the electric field is discontinuous across the circle, 
while the potential is continuous. 
Analogously to this problem, if Lee-Yang zeros accumulate on a curve in the thermodynamic limit, 
then the electric field $
\nabla_\xi \phi$ is discontinuous across the curve. 
As
$\nabla_\xi \phi$ is proportional to the 
(complexified) number density, 
the discontinuity in it signals
the first order phase transition
taking place across the curve.
}

\section{Lee-Yang zeros in QCD at high temperature}
\label{sec:jtheta}

\comment{In this section, we derive the Lee-Yang zeros of QCD. 
We start by recapitulating 
some features of QCD at pure imaginary chemical potential
and of its free energy at high temperature.
Then we calculate canonical partition functions by 
using the saddle point approximation,
leading to the analytical determination of Lee-Yang zeros.}

\subsection{QCD at pure imaginary chemical potential}
%
\comment{Pure SU($N_c$) Yang-Mills theory, such as quenched QCD, has 
a center symmetry 
$\mathbb{Z}_{N_c},$ where $N_c$ is the number of colors, and $N_c=3$ in QCD. 
A 
transition from the hadronic phase to the QGP phase occurs at high temperature. 
The Polyakov loop defined by
\begin{align}
L(\vec{x})= \frac13\,\tr\, {\rm P} \biggl\{ \exp\biggl(i g \oint_0^{1/T} A_4(\vec{x},\tau) d \tau \biggr)\biggr\}, 
\end{align}
is related to the free energy of static quarks and can be used 
as an order parameter of the deconfinement transition.
Here ${\rm P}$ denotes the path-ordering~\cite{Peskin:1995aaa}, 
$\vec{x}$ the spatial coordinate, 
$\tau$ the imaginary time, 
$g$ the gauge coupling constant of QCD, 
and $A_4$ the $\mu=4$ component of the gauge field $A_\mu$ in QCD. 
$A_\mu$ is an su(3) matrix, and the trace is taken over color indices. 
Since the Polyakov loop is $\mathbb{Z}_3$-variant, nonzero values of the Polyakov 
loop $\langle L \rangle$  mean that the center symmetry is spontaneously broken, where 
the bracket $\langle \cdots \rangle$ denotes the average over gauge fields. 
In the deconfinement phase, QCD has three degenerate vacua according to the center 
symmetry, and one of them is favored to break the $\mathbb{Z}_3$ symmetry spontaneously. 

In the presence of quarks, the center symmetry is explicitly broken. 
The deconfinement phase transition turns into a smooth crossover without 
discontinuity in thermodynamic quantities. 
Nevertheless, the Polyakov loop is still used to distinguish the confinement 
and deconfinement phases. 

Roberge and Weiss found~\cite{Roberge:1986mm} that even in the presence of the quarks, 
the grand canonical partition function of SU($N_c$) gauge theory is invariant under the shift of 
$\mu_I=\Im \mu$ as
\begin{align}
Z\biggl( \frac{\mu_I}{T} \biggr) = Z\biggl( \frac{\mu_I}{T} + \frac{2\pi}{N_c} \biggr). 
\label{Eq:2014Nov27Eq1}
\end{align}
This states that the grand canonical partition function 
is periodic with respect to $\mu_I/T$ with the period of $2\pi/N_c$. 
This is referred to as the Roberge-Weiss (RW) periodicity. 

Roberge and Weiss also found a first-order phase transition at 
$\mu_I/T= (2k+1) \pi/N_c, (k=1, 2, \cdots, N_c)$ under the increase of $\mu_I$ at high temperature~\cite{Roberge:1986mm}. 
The argument of the Polyakov loop $\omega = \arg\langle L\rangle$ is often used 
as an order parameter of the
phase transition. 
In the deconfinement phase and in the presence of quarks with $\mu=0$, 
$\omega$ takes zero in the ground state, while the other two 
$\mathbb{Z}_3$ vacua are local minima. 
As $\mu_I$ is increased at high temperature, 
the Polyakov loop jumps from one of 
the $\mathbb{Z}_3$ vacua 
to another at $\mu_I/T=\pi/3, \pi, 5\pi/3$, and
the argument of the Polyakov loop changes discontinuously;   
\begin{align}
\omega = \left\{ 
\begin{matrix} 
             0 &  (0 \le \frac{\mu_I}{T} \le \frac{\pi}{3}, \frac{5\pi}{3} \le \frac{\mu_I}{T} \le 2\pi), \\
\frac{4\pi}{3} & (\frac{\pi}{3} \le \frac{\mu_I}{T} \le \pi), \\
 \frac{2\pi}{3} & (\pi \le \frac{\mu_I}{T} \le \frac{5 \pi}{3}).
\end{matrix}
\right. 
\end{align}
This transition is referred to as the 
RW phase transition.  
Roberge and Weiss found
the presence of the RW phase transition by using a perturbative analysis, 
and its absence
at low temperature using a strong coupling analysis. 
This was later confirmed non-perturbatively in lattice QCD simulations 
for several different setups~
\cite{deForcrand:2002ci,D'Elia:2002gd,D'Elia:2004at,D'Elia:2007ke,D'Elia:2009tm,Cea:2010md,deForcrand:2010he,Nagata:2011yf,Cea:2012ev,Bonati:2014kpa}.
It was also studied in effective models, see e.g.~\cite{Kashiwa:2013rm,Sakai:2008py,Morita:2011jva}.

A nonzero value of $\omega$ means that the gauge field $A_4$ acquires 
an expectation value $\omega T/g$.
As we will show in the next subsection, this effect plays a 
crucial role in realizing 
the RW periodicity of the free energy at high temperature. 
In this work, we assume that the system is homogeneous and in equilibrium.
We also assume a priori that 
the thermal fluctuation of $\omega$ and $A_4$ is negligibly small. 
Under this assumption we fix $A_4$ at its classical value
in a background field method~\cite{Peskin:1995aaa}. }

\comment{Using the RW periodicity,
canonical partition functions $Z_n$ are classified in terms of 
the value of $n \; \mod \; N_c$, which is referred to as the triality
for QCD.
We also refer $\{n| n\equiv 0\; \mod \; 3 \}$ as the triality sector, 
and the other two sectors, $\{n| n\not\equiv 0 \;\mod \;
3 \} $ as the non-zero triality sector.
Using Eqs.~(\ref{Eq:2014Mar16eq0}) and (\ref{Eq:2014Nov27Eq1}), we can show
~\cite{Kratochvila:2006jx} that 
\begin{align}
Z_n = 0,\ n \not\equiv 0
\; \mod\  3.
\label{Eq:2014Jan08eq1}
\end{align}
This means that only the triality sector contributes to $Z(\mu)$ due to the RW periodicity, 
while the nonzero triality sector does not. 
Using Eq.~(\ref{Eq:2014Jan08eq1}), the fugacity polynomial Eq.~(\ref{Eq:2014Mar16eq0}) is expressed 
in terms of the baryon number instead of the quark number as 
\begin{align}
Z(\mu) &= \sum_{
n \equiv 0 \; \mod \; 3 } Z_n \xi^n
       = \sum_{n_{\rm B}} Z_{n_{\rm B}} \xi_{\rm B}^n, 
\label{Eq:2014Dec31eq1}
\end{align}
where $\xi_{\rm B}=\exp(3 \mu/T)=\xi^3$ and $n_{\rm B} = n/3$. 
We rewrite the canonical partition functions $Z_n$ for quark number $n=3n_{\rm B}$ 
as those for baryon number $Z_{n_{\rm B}}$ to obtain the right hand side. 
}

\subsection{Free energy at temperatures above $T_c$}

At high temperature, the free energy density of QCD 
eventually approaches a quartic polynomial with even powers of $\mu/T$~\cite{Kapusta:2006aaa,Allton:2003vx}:
\begin{align}
- \frac{f(\mu)}{T^4} = c_0  + c_2 \biggl( \frac{\mu}{T}\biggr)^{2}
+ c_4 \biggl( \frac{\mu}{T}\biggr)^{4}.  
\label{Eq:2014Apr24eq1}
\end{align}
The minus sign is conventionally introduced.

\comment{
The question arises as to whether higher order coefficients survive
in cases other than the Stefan-Boltzmann (SB) limit,
such as in the presence of interaction
or at moderate temperature, etc. 
Lattice QCD simulations suggested that the free energy indeed approaches
Eq.~(\ref{Eq:2014Apr24eq1}) at temperature slightly 
above the pseudo critical temperature $T_c$. 
The sixth order term $c_6$ has been calculated in lattice QCD with 
different setups: the p4-improved staggered fermions with the bare quark mass $m/T=0.4$ 
on a $16^3\times 4$ lattice~\cite{Allton:2005gk}, 
at two different pion masses $m_\pi=220$ and $770$ MeV\cite{Miao:2008sz}, 
the clover-improved Wilson fermions on $8^3\times 4$ with 
the pion mass about $800-1000$ MeV~\cite{Nagata:2012pc}.
A common feature is that $c_6$ rapidly decreases with temperature for $T>T_c$ and 
vanishes at a certain temperature. 
The vanishing temperature of $c_6$ was estimated to be 
$T=(1.1-1.2)T_c$  in~\cite{Allton:2005gk,Miao:2008sz,Nagata:2012pc}. 
Usually, Taylor coefficients are calculated using 
the so-called noise method~\cite{Ejiri:2009hq}, 
which associates with a randomly generated vector. 
This method becomes less useful
for higher order terms due to the numerical 
uncertainty caused by the random noise vector. 
On the other hand, in~\cite{Nagata:2012pc}, we used a reduction formula~\cite{Nagata:2010xi}. 
The reduction formula provides a method to evaluate the Taylor coefficients 
without the random noise vector, although
its applicability has been limited to small lattice volumes.
Using the formula, 
the Taylor coefficients was calculated up to tenth order, 
and the values of $c_8$ and $c_{10}$ were indeed
consistent with zero at $T>1.1 T_c$ within error bars. 
Thus, the lattice QCD simulations  are unanimous in supporting
\cite{Allton:2005gk,Miao:2008sz,Nagata:2012pc} that 
the free energy takes the quartic form at temperature 
$(1.1-1.2) T_c$ or above.} 

\begin{figure}[htbp] 
\includegraphics[width=7cm]{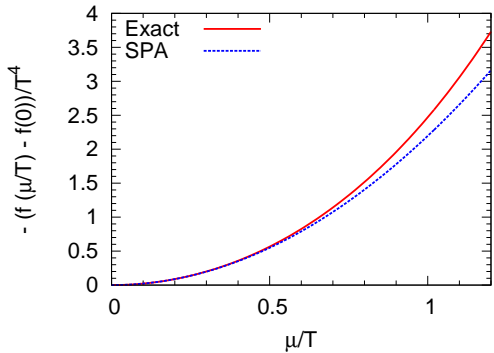}
\caption{Comparison of the saddle point approximation (SPA) and exact result 
for the free energy density. We use values of $c_2$, $c_4$ and $VT^3$ used 
in our simulations in the next section ($c_2=2.20$, $c_4=0.27$\comment{).}}
\label{Fig:2014May29fig1}
\end{figure}

\comment{
Lattice QCD simulations also showed that $c_2$ is 
larger compared to $c_4$ at high temperature. 
Below, we will use a saddle point approximation, which requires $c_2$ 
to be sufficiently large compared to $c_4$. 
Here, we estimate the validity range of this approximation. 
$c_2$ and $c_4$ have been well studied in lattice QCD simulations, 
e.g.~\cite{Allton:2005gk,Miao:2008sz,Ejiri:2009hq,Nagata:2012pc}. They are comparable 
in magnitude at $T\sim T_c$, and 
$c_2$ ($c_4$) increases (decreases) monotonically 
and rapidly as the temperature is raised above $T_c$,
so that $c_2$ surpasses $c_4$
at temperature to some extent higher than $T_c$. 
The lattice QCD simulations reported ~\cite{Allton:2005gk,Miao:2008sz,Nagata:2012pc,Ejiri:2009hq} 
that $c_2$ is about 10 times larger than $c_4$ at temperatures
$T=(1.1-1.2) T_c$.
As we will see below, this
justifies the use of the saddle point approximation. 
Note that $c_2$ and $c_4$ 
deviate from those at the SB limit in lattice QCD simulations 
due to lattice artifacts. 
The discussion below is valid if the free energy 
satisfies the above two conditions, 
$c_6=c_8=\cdots=0$ and $c_2\gg c_4$, 
regardless of whether 
the SB limit is reached or not. 

\subsection{Canonical partition functions and Lee-Yang zeros}
Having explained two features of the free energy of QCD at high 
temperature,
we limit our discussion to the case where these conditions hold
and 
study the canonical partition functions and Lee-Yang zeros.
First, we extract $Z_n$ by substituting Eq.~(\ref{Eq:2014Apr24eq1}) into  Eq.~(\ref{Eq:2014Dec04eq1}).
However, the free energy (\ref{Eq:2014Apr24eq1}) is obtained for $\Im \mu=0$. 
As we have remarked,
the argument of the Polyakov loop $\omega$, i.e. the field $A_4$,
acquires a nonzero expectation
value via the RW phase transition as $\mu_I$ is increased.
}
We assume $A_4$ to be constant, and absorb this 
condensate into the shift of the imaginary part of the
chemical potential : $\mu + i g A_4 = \mu +i \omega T$. 
By taking into account the contributions from \comment{the} three domains, 
$Z_n$ is given as 
\begin{align}
Z_n &= \int_{-\pi/3}^{\pi/3} e^{-V f(\theta)/T} e^{in\theta} \frac{d \theta}{2\pi} + \int_{\pi/3}^{\pi} 
e^{-V f(\theta - 2\pi/3)/T} e^{in\theta} \frac{d \theta}{2\pi} \nonumber \\
& + \int_{\pi}^{5\pi/3} e^{-V f(\theta-4\pi/3)/T} e^{in\theta} \frac{d\theta}{2\pi} , 
\end{align}
where $\theta = \mu_I/T$. Using the RW periodicity~\cite{Roberge:1986mm}, this reads
\begin{align}
Z_n = \int_{-\pi/3}^{\pi/3} e^{-V f(\theta)/T} e^{in\theta} (1 + e^{ -i  2 \pi n /3} + e^{ -i  4 \pi n / 3}) 
\frac{d\theta}{2\pi}. 
\label{Eq:2014Apr22eq1}
\end{align}
This ensures $Z_n = 0$ for $n \not\equiv 0 \ \mod\ 3$,
so that Eq.~(\ref{Eq:2014Apr22eq1}) is expressed as
\begin{align}
Z_n = \frac{3}{2\pi} \int_{-\pi/3}^{\pi/3} d \theta \; e^{T^3 V g(\theta) + i n\theta}\ \ (n \equiv 0 \ \mod\ 3).
\label{Eq:2014Mar16eq3}
\end{align}
where $g(\theta) = c_0 - c_2 \theta^2 + c_4 \theta^4$. 
The function $g(\theta)$ has one maximum at $\theta=0$ and two minima at $\theta = \pm \sqrt{c_2/(2 c_4)}$. 
\comment{
At high temperature, those two minima are outside of the 
integration domain. 
As we have mentioned above, $c_2$ and $c_4$ are comparable at $T=T_c$
and $c_2$ rapidly increases and $c_4$ rapidly decreases above $T_c$.
For instance, $\sqrt{c_2/(2 c_4)} = \sqrt{2}\pi$ in the SB limit~\cite{Kapusta:2006aaa,Allton:2003vx,Ejiri:2009hq}
and $\sqrt{c_2/(2 c_4)} \sim \sqrt{5}$ for $T=(1.1-1.2)T_c$
\cite{Allton:2005gk,Miao:2008sz,Nagata:2012pc,Ejiri:2009hq}.
$g(\theta)$ is a concave function for $\theta \in [-\pi/3, \pi/3]$ with the maximum 
at $\theta=0$.}
It has a sharper peak for larger volume, and the integral 
in Eq.~(\ref{Eq:2014Mar16eq3}) is dominated by $\theta=0$ for large $V$.
This allows for the use of the saddle point approximation 
\commentb{(SPA)} to Eq.~(\ref{Eq:2014Mar16eq3}) so that $Z_n$ is reduced to 
\begin{align}
Z_n = C e^{ - n^2/(4T^3V c_2)} \ \  (n\equiv 0\ \mod\ 3), 
\label{Eq:2014Apr21eq1}
\end{align}
where $C= \frac{3}{2\pi} \sqrt{\frac{\pi}{T^3 V c_2}} e^{T^3 V c_0}$. 
\commentb{The quartic term in $g(\theta)$ vanishes according to the saddle point approximation.}
In transition from Eq.~(\ref{Eq:2014Mar16eq3}) to (\ref{Eq:2014Apr21eq1}), we approximated
an incomplete gamma function by a complete counterpart (see the Appendix).
The validity of our approximation is estimated by the condition
$c_2 (\mu/T)^2 \gg c_4 (\mu/T)^4$ so that the free energy is dominated by the second-order term.
To evaluate the applicable range of Eq.~(\ref{Eq:2014Apr21eq2}), 
we have plotted
$(f(\mu)-f(0))/T^4$ for the saddle point approximation 
and the original one (\ref{Eq:2014Apr24eq1})
in Fig.~\ref{Fig:2014May29fig1}.
It indicates that the
saddle point approximation is valid for small $\mu/T$ $\lesssim 0.5.$

Assuming the validity of Eq.~(\ref{Eq:2014Apr21eq1}), $Z(\mu)$ is reconstructed as 
\begin{align}
Z(\mu) &= C \sum_{{n_{\rm B}}=-\infty}^\infty e^{- 9 {n_{\rm B}}^2/(4T^3 V c_2) + 3 {n_{\rm B}} \mu/T}. 
\label{Eq:2014Apr21eq2}
\end{align}
The zeros of the grand canonical partition function are readily obtained
by recognizing that Eq.~(\ref{Eq:2014Apr21eq2}) is equal to the Jacobi 
theta function $\vartheta(z, \tau) = \sum_{n=-\infty}^{\infty} e^{\pi i n^2 \tau + 2 \pi i n z}$,
\begin{align}
Z(\mu) \propto \vartheta(z, \tau), 
\end{align}
where $z$ and $\tau$ are given by 
\begin{align}
2\pi i z  = 3 \frac{\mu}{T},
\quad
\pi i \tau = - \frac{9}{4T^3 V c_2}.
\end{align}
Thus the Lee-Yang zeros of Eq.~(\ref{Eq:2014Apr21eq2}) are given by the zeros 
of $\vartheta(z,\tau)$ located at
\begin{align}
\frac{\mu}{T} = \frac{(2k+1) \pi i}{3} - \frac{3(2 \ell +1)}{4 T^3 V c_2},
\label{Eq:2014Apr28eq1}
\end{align}
where $k$ and $\ell$ take all integer values
as a consequence of the pseudodouble periodicity of the theta function.

On the complex plane of the baryon fugacity $\xi_{\rm B} = \xi^3$, 
all zeros in Eq.~(\ref{Eq:2014Apr28eq1}) are located on the negative real axis. 
On the complex $\xi$ plane, Eq.~(\ref{Eq:2014Apr28eq1}) 
lies on three radial lines at arguments
$\arg \xi =\Im (\mu/T) = \pi/3, \pi,$ and $5\pi/3$. 
The RW phase transition occurs at the points
$(\Re(\mu/T), \Im(\mu/T)) =(0, (2k+1)\pi/3)$~\cite{Roberge:1986mm}. 
The Lee-Yang zeros closest to these points are given by 
\begin{align}
\frac{\mu}{T} = \frac{(2k+1) \pi i}{3} \pm \frac{3}{4 T^3 V c_2}. 
\label{Eq:2014Mar22eq1}
\end{align}
Each of them approaches the corresponding RW phase transition point 
in \comment{the} thermodynamic limit as $1/V$. 
This explains the first-order nature of the RW phase transition according 
to the Lee-Yang zero theorem. 
In addition, Eq.~(\ref{Eq:2014Apr28eq1}) also indicates that the RW phase transition 
occurs at $\mu_I/T = \pi/3$ even for $\re(\mu/T)\neq 0$,
as long as the saddle point approximation is valid.
We note that it is possible to obtain the Lee-Yang zeros of Eq.~(\ref{Eq:2014Mar22eq1}) 
directly from the free energy by using a method \comment{presented} in~\cite{Biskup:2000prl1}.

\section{Lattice QCD simulations}
\label{sec:simulation}
\subsection{Method and setup}
In this section, we reexamine the data of our previous lattice QCD simulations,
in which three quantities, the RW phase transition~\cite{Nagata:2011yf},
Taylor coefficients of the free energy~\cite{Nagata:2012pc},
canonical partition functions and Lee-Yang zeros~\cite{Nagata:2012tc,Nakamura:2013ska},
were calculated in the same lattice setup.
Below we recapitulate the calculation of $Z_n$ and Lee-Yang zeros on the lattice,
and summarize the setup of the simulations to make the paper self-contained.

The grand canonical partition function of lattice QCD is given by
\begin{align}
Z(\mu) = \int {\cal D} U (\det \Delta(\mu))^{N_f} e^{-S_g},
\label{Eq:2014Apr22eq2}
\end{align}
where $U$, $\Delta(\mu)$, and $S_g$ denote link variables, fermion matrix, and gauge action, respectively.
We employ a clover-improved Wilson fermion action with $N_f=2$ and renormalization-group 
improved gauge action~\cite{AliKhan:2000iz}.

We calculate $Z_n$ using a Glasgow method~\cite{Barbour:1991vs,Hasenfratz:1991ax}.
We expand the fermion determinant in powers of $\xi$ using a reduction formula of the Wilson fermion 
determinant~\cite{Nagata:2010xi,Gibbs:1986hi,Hasenfratz:1991ax,Adams:2003rm,Borici:2004bq,Alexandru:2010yb}, 
\begin{align}
(\det \Delta (\mu))^{N_f} = \sum_{n=-2N_f N_s^3 }^{2 N_f N_s^3} d_n \xi^n, 
\label{Eq:2014May03eq1}
\end{align}
which provides the fugacity expansion of $Z(\mu)$. 
Since $d_n$ is complex, it is not possible to use $d_n$ as a measure 
for Monte Carlo simulations. 
Instead we use Ferrenberg-Swendsen reweighting for the fermion determinant\comment{: }
$\det \Delta(\mu) = (\det \Delta(\mu)/\det \Delta(0)) \det \Delta(0)$
and express $Z(\mu)$ as an expectation value of the operator 
$\det \Delta(\mu)/\det \Delta(0)$ averaged over gauge configurations generated at $\mu=0$. 
Then $Z_n$ is given by 
\begin{align}
Z_n  & = \int {\cal D}U \frac{d_n}{(\det \Delta(0))^{N_f}} (\det \Delta(0))^{N_f} e^{-S_g}, \nonumber \\
     & = Z_0 \left\langle \frac{d_n}{(\det \Delta(0))^{N_f}} \right\rangle_0, 
\label{Eq:2014May14eq1}
\end{align}
where $Z_0 = \int {\cal D}U (\det \Delta(0))^{N_f} e^{-S_g}$, and 
$\langle \cdots \rangle_0$ denotes the expectation value obtained from 
gauge configurations generated at $\mu=0$ with reweighting. 

The simulation was performed in the following setup\comment{:}
The lattice volumes were $N_s^3 \times N_t = 8^3 \times 4$ and $10^3\times 4$ with 
spatial and temporal lattice sizes  $N_s$ and $N_t$. 
The simulation was performed along the line of constant physics with 
$m_{\pi}/m_{\rho} = 0.8$~\cite{Ejiri:2009hq}.
We considered two temperatures $T/T_c=0.99$\; ($\beta=1.85$) and $1.20\;(1.95)$, 
where $\beta =6/g^2$ is the bare lattice coupling constant, and $T_c$ is 
the pseudocritical temperature at $\mu=0$. 
The RW phase transition point was estimated to be at $\beta=1.92$~\cite{Nagata:2011yf}
and 11 000 HMC trajectories were simulated for each parameter set. 
The observables were calculated using 400 configurations with 20-trajectory intervals
after removing the initial 3000 trajectories for thermalization.  

Lee-Yang zeros are obtained by using a method based on the Cauchy integral theorem 
with a divide-and-conquer algorithm and multiprecision arithmetic~\cite{Nakamura:2013ska,web:FMlib}. 

\comment{
We apply the RW periodicity to $Z(\mu)$ and express it in terms of the baryon number, 
as shown in Eq.~(\ref{Eq:2014Dec31eq1}). 
We solve $Z(\mu)=0$ for Eq.~(\ref{Eq:2014Dec31eq1}), and obtain Lee-Yang zeros for $\xi_{\rm B}$. 
They are transformed into
the zeros for $\xi$ by using $\xi=\xi_{\rm B}^{1/3}$. 
Thus, obtained Lee-Yang zeros automatically satisfy 
the $\mathbb{Z}_3$ symmetry on the complex $\xi$ plane. 
}
For further detail, see \cite{Nagata:2012pc,Nagata:2012tc,Nakamura:2013ska}.

\subsection{Canonical partition functions}
\begin{figure}[htbp] 
\includegraphics[width=8cm]{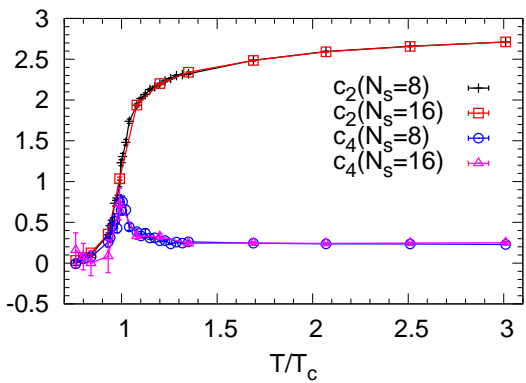}
\caption{Second- and fourth-order coefficients of Taylor expansion of 
the free energy, $c_2$ and $c_4$, for $N_s=8$ and $N_s=16$. 
The data for $N_s=8$ and $N_s=16$ are  taken from \cite{Nagata:2012pc} and \cite{Ejiri:2009hq}, respectively.}
\label{Fig:2014Mar17fig1}
\end{figure} 
The $T$ and $V$ dependences of Taylor coefficients
$c_2$ and $c_4$ are plotted in Fig.~\ref{Fig:2014Mar17fig1}.
At $T/T_c=0.99$, $c_2$ and $c_4$ are comparable in magnitude, 
while $c_2$ is several times larger than $c_4$ at $T/T_c=1.20$. 
We observed in ~\cite{Nagata:2012pc} that higher-order coefficients $c_6, c_8,$ 
and $c_{10}$ are consistent with zero at $T/T_c=1.20$ so that $f(\mu)$
approaches the quartic function of $\mu$ as expected in (\ref{Eq:2014Apr24eq1}). 
We also observe that the coefficients $c_2$ and $c_4$ are insensitive to 
the lattice volume.  
Thus, the conditions used in the saddle point approximation are satisfied 
at $T/T_c=1.20$ or higher temperatures. 

\begin{figure}[htbp] 
\includegraphics[width=8cm]{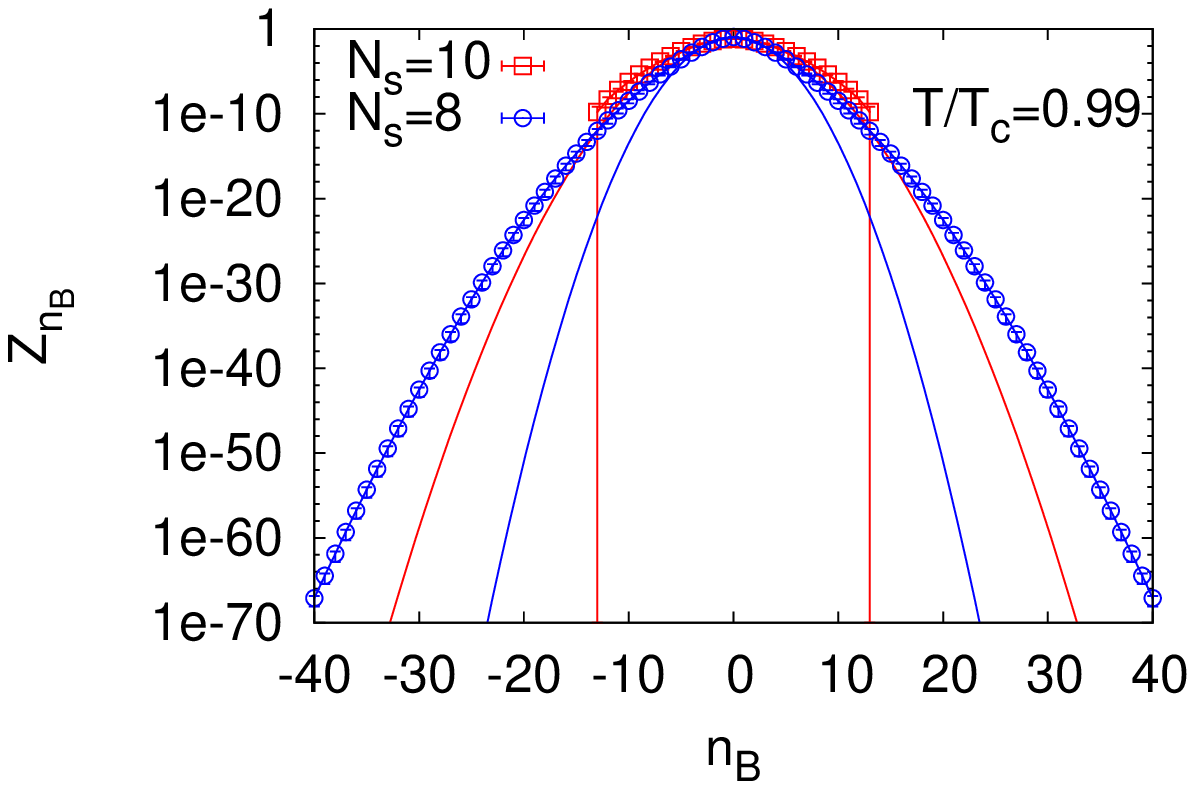}
\includegraphics[width=8cm]{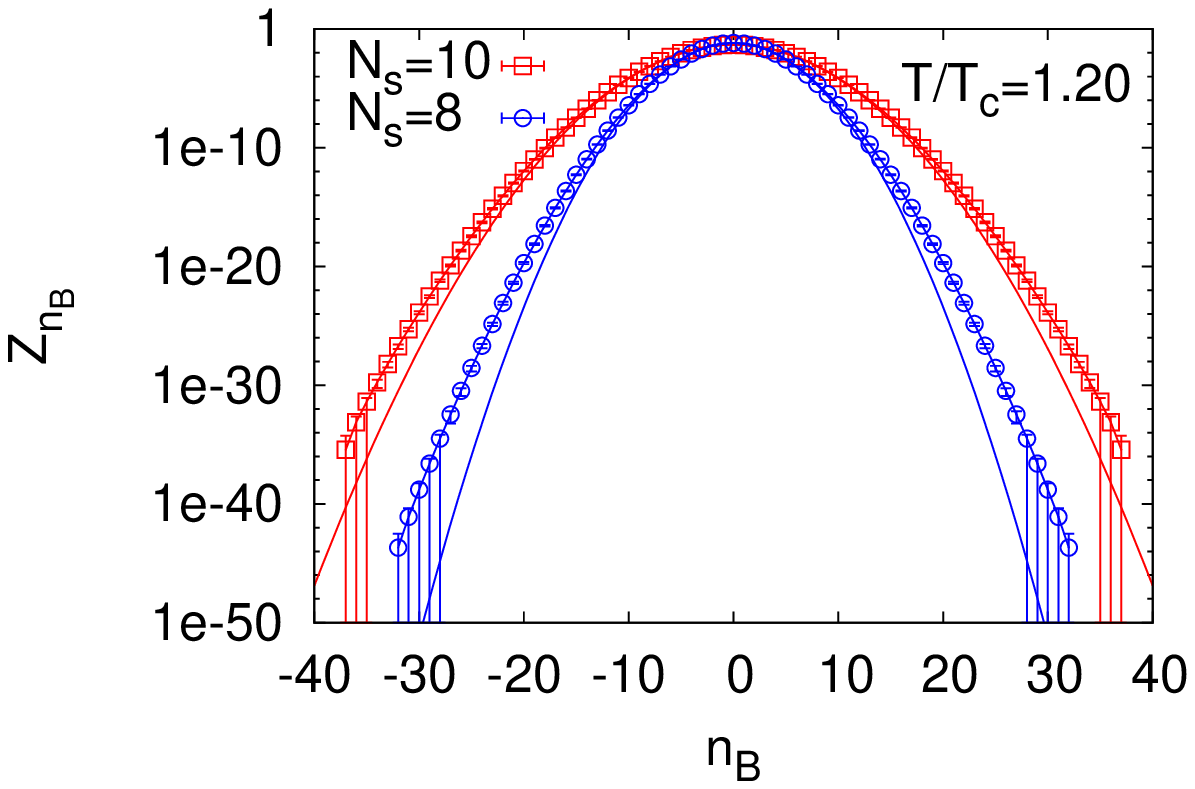}
\caption{Canonical partition function as a function of the 
baryon number for $T/T_c=0.99$ (top) and $1.20$ (bottom).
The data are obtained from the canonical formalism, 
while the solid curves are obtained from the saddle point approximation. 
Note that only positive $Z_n$'s are shown. 
}
\label{Fig:2014Mar17fig2}
\end{figure} 
We plot the canonical partition functions $Z_{n_{\rm B}}$ ($n=3n_{\rm B}$) in Fig.~\ref{Fig:2014Mar17fig2}.
The squares and circles indicate the values obtained from the canonical approach,
Eq.~(\ref{Eq:2014May14eq1}). 
Since the average of $Z_n$ can be negative for large $n$ due to the overlap problem,
we plot the values of $Z_{n_{\rm B}}$ up to $\pm n_{\rm B}$ below which 
the partition functions are all positive.
The solid curves represent the Gaussian functions (\ref{Eq:2014Apr21eq1}) 
with $c_2$ obtained from the lattice simulation. 
We observe that $Z_{n_{\rm B}}$ with relatively small $n_{\rm B}$
follows the Gaussian function at 
$T/T_c=1.20$ as expected, while they fail to match at $T/T_c=0.99$. 

\begin{figure}[htbp] 
\includegraphics[width=8cm]{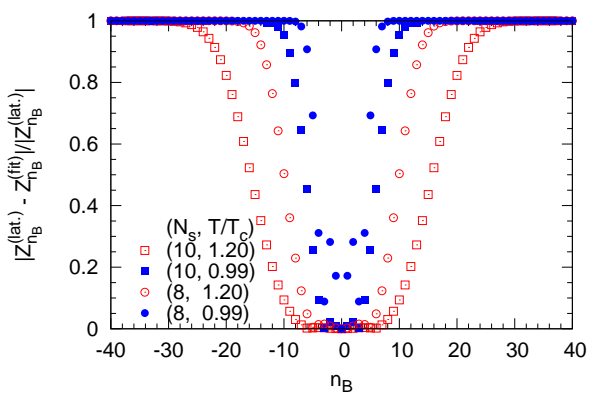}
\caption{The difference of the canonical partition functions 
between the data $(Z_{n_{\rm B}}^{(\rm lat)})$ and Gaussian fit $(Z_{n_{\rm B}}^{(\rm fit)})$. 
Circle and square symbols denote results for $N_s=8$ and $10$, respectively.
Open (red) and closed (blue) symbols are for $T/T_c=1.20$ and $0.99$, respectively.
}
\label{Fig:2014Sep03fig1}
\end{figure} 
In order to quantify the consistency, we plot the relative difference between the lattice data and 
Gaussian fit in Fig.~\ref{Fig:2014Sep03fig1}. 
The data and Gaussian functions show better agreement for higher temperature and larger volume. 
However, the Gaussian function systematically deviates from the data for large $n$ even at high $T$.
This deviation is partly caused by the smallness of the lattice volume,
as a better agreement is observed for larger volume. 
It is likely that the deviation may originate from the breakdown of the saddle point approximation, 
as its validity is limited to small $\mu/T$. 
Below, we shall examine how the deviation affects the distribution of Lee-Yang zeros.
\comment{Kratochvila and de Forcrand showed the agreement of the free energy obtained from the canonical approach 
and Taylor expansion using a staggered fermion action~\cite{Kratochvila:2005mk,Kratochvila:2006jx}. }

\subsection{Lee-Yang zeros}
In this work we calculate Lee-Yang zeros for $T/T_c=1.20$. 
Before proceeding to numerical results, we remark on the
numerical instability of the canonical partition functions at large $n$. 
The fugacity coefficients $d_n$ take complex values for each configuration. 
The phase of $d_n$ fluctuates more rapidly for larger $n$, because its 
modulus is exponentially suppressed as $n$ is increased.
Beyond a certain value of $n$, $Z_n$ becomes negative, 
and the inclusion of such $Z_n$ would yield unphysical zeros of $Z(\mu)$ and 
cause unphysical nonanalyticity for the free energy, even in a finite volume. 
Accordingly, we are obliged to truncate the fugacity polynomial at 
$|n_{\rm B}|=n_0$ so that all $Z_{n_{\rm B}}$'s are positive for 
$|n_{\rm B}| \le n_0$.
\comment{In the following we will consider three different cases 
of truncation:
(a) 
$n_0=37$ and $N_s=10$, (b) $N_s=8$ and $n_0=32$ and (c) 
$N_s=8$ and $n_0=19$. 
For a larger spatial lattice of size $N_s=10$,
it is natural to take
(a) the maximal permissible value $n_0=37$ 
as seen from Fig.~\ref{Fig:2014Mar17fig2}.
For a smaller spatial lattice $N_s=8$,
we try the following two alternatives:
(b) maximal permissible  value $n_0=32$ as seen from Fig.~\ref{Fig:2014Mar17fig2}, and
(c) $n_0=19\simeq 37\times(8/10)^3$ so that the truncation order is proportional to the lattice volume
as compared to the case with $N_s=10$.
} 

Below, we examine the convergence of the fugacity polynomial by comparing these two choices.
\begin{figure}[htbp] 
\includegraphics[width=8cm]{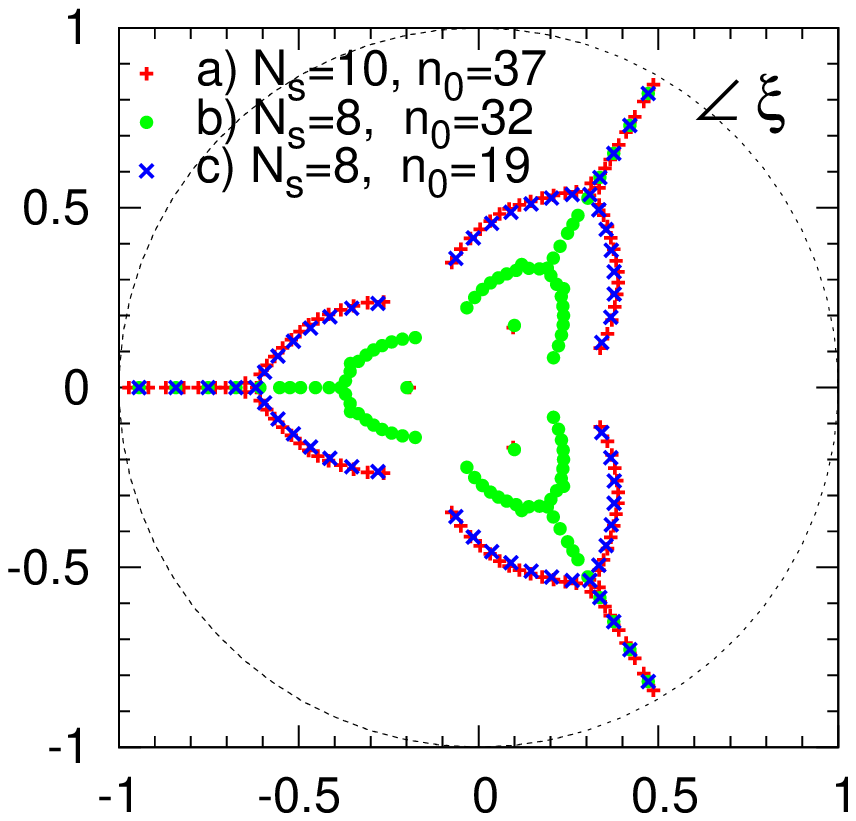}
\includegraphics[width=8cm]{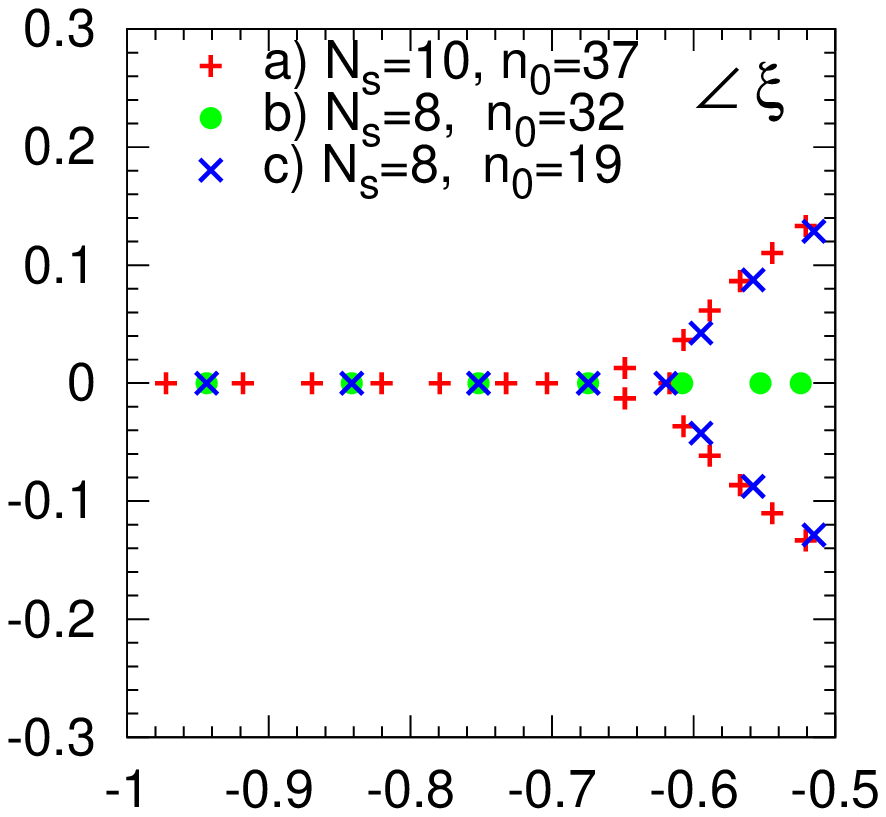}
\caption{Lee-Yang zeros on the complex fugacity plane for $\beta=1.95, T/T_c=1.20$.
Top panel: Zeros inside unit circle. Bottom panel: Zeros on 
or in the vicinity of the negative real axis. 
Red, green, and blue symbols denote Lee-Yang zeros for 
(a) $n_0=37$ and $N_s=10$, (b) $N_s=8$ and $n_0=32$, and (c) 
$N_s=8$ and $n_0=19$, respectively. 
Note that zeros also exist outside the unit circle with symmetry $\xi\leftrightarrow 1/\xi$.}
\label{Fig:2014Mar17fig3}
\end{figure} 
Figure \ref{Fig:2014Mar17fig3} shows the distributions of the Lee-Yang zeros on the complex plane of the quark fugacity $\xi$,
corresponding to the cases (a) red, (b) green, and (c) blue, respectively.
The distributions on the baryon fugacity plane are readily obtained by
using the relation $\xi_{\rm B}=\xi^3$. 
Near the unit circle, the Lee-Yang zeros are located on three
radial lines with arguments $\arg \xi  = \pi/3, \pi,$ and $5\pi/3$. 
This behavior is qualitatively consistent with the prediction in Eq.~(\ref{Eq:2014Apr28eq1})
\commentb{and indicative of the RW phase transition.}
As the origin is approached, each line branches to two curves. 
\comment{Barbour et
al.~\cite{Barbour:1991vs} obtained 
this behavior of the Lee-Yang zeros in
their pioneering study of finite density lattice QCD simulations. 
Specifically they found the zeros on the twelve radial lines on the 
$e^{\mu}$-plane. 
In this work, we take a further step 
to confirm this interpretation by examining 
the volume scaling and the asymptotic convergence, and 
by comparing them with the analytic calculation.}
The zeros near the unit circle $(0.6 < |\xi| <1)$ are stable
as $n_0$ is increased from 19 to 32 for $N_s=8$, 
which indicates the convergence of the fugacity polynomial. 
On the other hand, the increment of $n_0$ affects the location of the zeros for large chemical potential (small $\xi$). 
We also observe that under a shift of $n_0$ as proportional to the volume 
($n_0=19$ for $N_s=8$ and $n_0=37$ for $N_s=10$), the Lee-Yang zeros are located on the 
common trajectories, and the density of the zeros are doubled. 

\begin{figure}[htbp] 
\includegraphics[width=8cm]{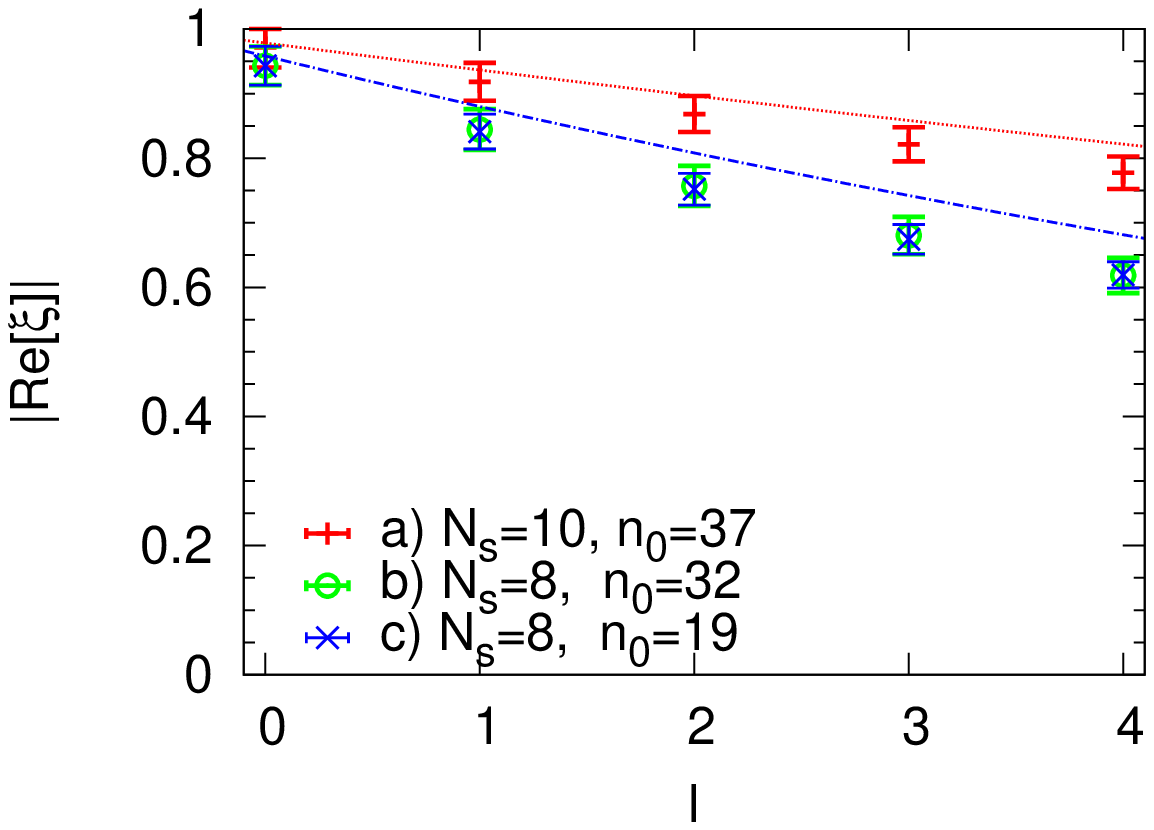}
\includegraphics[width=8cm]{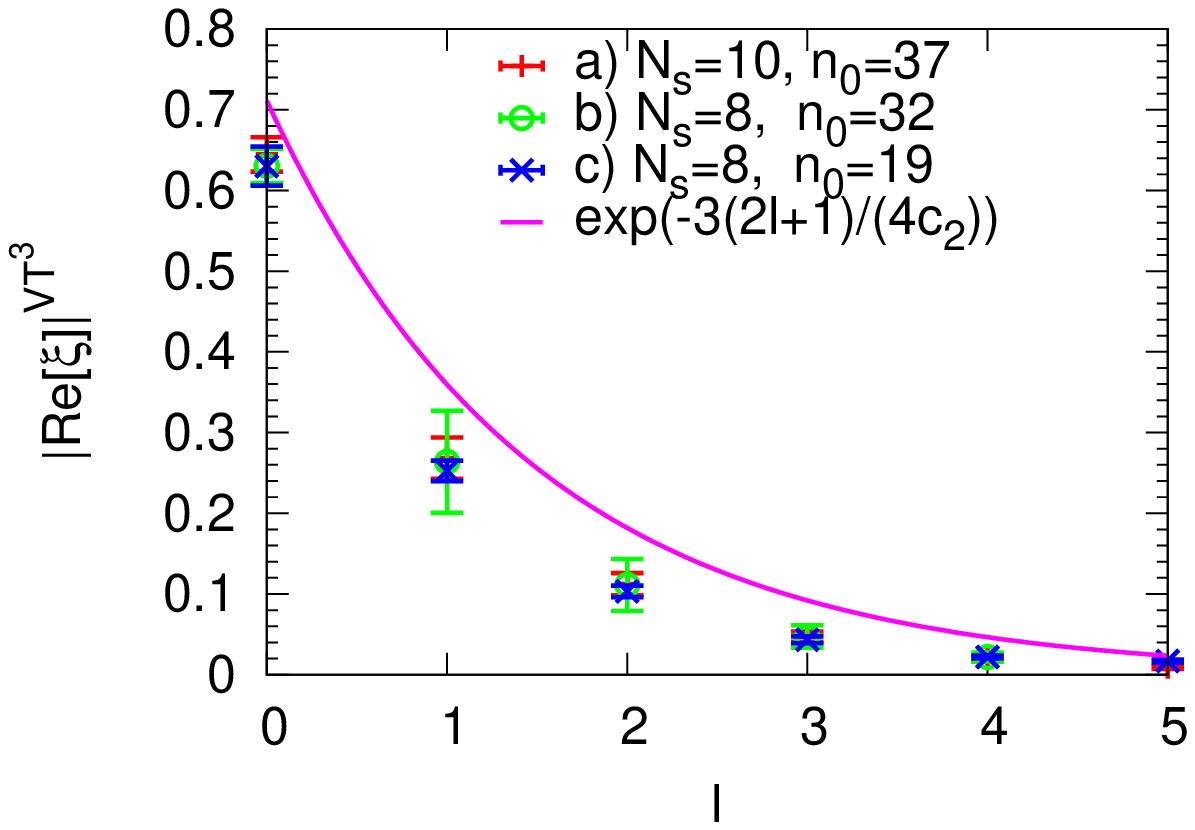}
\caption{Top: $|\Re\, \xi|$ for zeros on the negative real axis near the unit circle, 
\comment{where $\xi=\exp(\mu/T)$.}
Red, green, and blue symbols denote Lee-Yang zeros for 
(a) $n_0=37$ and $N_s=10$, (b) $N_s=8$ and $n_0=32$, and (c) 
$N_s=8$ and $n_0=19$, respectively. 
The curves represent predictions from the Jacobi theta function
$\exp(-3(2\ell+1)/ (4T^3V c_2))$ for $N_s=10$ (red) and $N_s=8$ (green and blue).
Bottom: Volume-independent combination $|\Re \, \xi|^{VT^3}$.
The curve represents $\exp(-3(2\ell+1)/ (4 c_2))$.}
\label{Fig:2014Sep03fig2}
\end{figure} 
In order to make a quantitative comparison between lattice and analytic results, 
in Fig.~\ref{Fig:2014Sep03fig2} we plot $|\re\, \xi|$ (top) and $|\re\, \xi|^{VT^3}$ (bottom)
for several Lee-Yang zeros near the unit circle on the negative real axis. 
Statistical errors of the Lee-Yang zeros in the plot are estimated
with a bootstrap method as follows\comment{: }
For each bootstrap sample, we calculate $Z_n$ up to $n_0$ and locate the Lee-Yang zeros. 
Since we are interested in the zeros relevant to the RW phase transition, 
we pick up some zeros near the unit circle
and label them as $\ell=1, 2, \ldots$ in the order of modulus.
For each label $\ell$,  statistical errors are estimated
as the variance over 1000 bootstrap samples.
Note that the zeros at $|\xi| \lesssim 1$ shown
in Fig.~\ref{Fig:2014Sep03fig2} indicate no 
fluctuation in the imaginary part, while the zeros with smaller $|\xi|<0.6$
fluctuate both in their real and imaginary parts. 
We observe that each Lee-Yang zero calculated in the simulation is
systematically smaller in magnitude than the zero of the corresponding order 
predicted in the saddle point approximation. 
In principle, there could be two possible origins for this deviation:
slow convergence of the fugacity expansion (\ref{Eq:2014May03eq1}) and/or
deviation of the canonical partition functions from the Gaussian function. 
As there is no systematic difference between 
the two choices of the truncation order $n_0=19$ and $32$, 
we can exclude the former origin and safely conclude that 
the deviation is caused by the deviation of $Z_n$ 
in the large-$n$ sector from the Gaussian function.
Despite this systematic deviation, the saddle point approximation well explains the features
of the lattice data, such as trajectory of zeros, spacing between zeros, and volume dependence. 

\begin{figure*}[htb] 
\includegraphics[width=13cm]{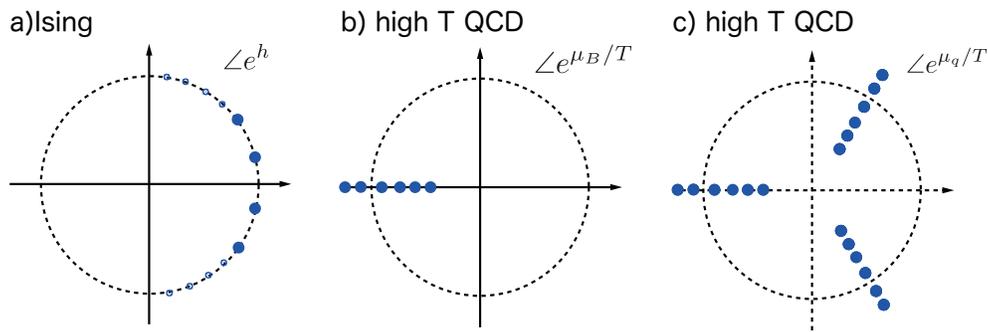}
\caption{Schematic figures for the distribution of Lee-Yang zeros in
several cases. (a) Ising models on the complex plane for $e^{h}$, where 
$h$ is the external magnetic field in Ising models, 
(b) QCD on the complex plane for baryon fugacity, 
and (c) QCD on the complex quark fugacity plane. Dotted circles denote
the unit circle. 
Case (b) can be generalized to free fermion theories.}
\label{Fig:2014Sep21fig1}
\end{figure*} 
The result is indicative in the viewpoint of the Lee-Yang zero theorem. 
As we have discussed, if the canonical partition function is Gaussian, 
then zeros are located on the negative real axis (for the baryon fugacity). 
Thus, theories with the Gaussian type of canonical partition functions, such as 
a gas of free fermions at small chemical potential, are 
exceptional cases of the Lee-Yang zero circle theorem (Fig.\ref{Fig:2014Sep21fig1}). 
QCD is expected to be an exceptional case of the Lee-Yang zero circle theorem if it undergoes 
a phase transition at $\Re\,\mu \neq 0$. 
This can be trivially proven for the case with even number of flavors: Then, the
Boltzmann weight is real and positive on the unit circle on the complex fugacity plane,
and zeros cannot exist on the unit circle. 
A concrete example of the Lee-Yang zeros provided in the present work
would help to deepen our understanding of the Lee-Yang zero theorem. 

\comment{
We obtained the Lee-Yang zeros as the roots of the fugacity polynomial. 
An alternative method 
would be to calculate zeros of the grand partition function
with a reweighting method. 
However, the validity of the latter is controversial because 
Lee-Yang zeros appear when the average of a reweighting factor vanishes. 
This seems to imply the breakdown of the reweighting method. 
Moreover this problem tends to become severer
for 
\comment{larger} volume,
which makes it difficult to distinguish physical zeros from
contamination of the 
sign problem~\cite{Ejiri:2005ts}.
%
The same problem does indeed occur even in the canonical approach; 
the sign problem sometimes makes $Z_n$ negative, which allows $Z(\mu)$ to vanish 
for real quark chemical potentials even in a finite volume.
Such zeros are unphysical ones 
originated from the sign problem.
In this work, we have circumvented this problem by the truncation of the fugacity 
polynomial so that all $Z_n$ are positive. 
Although the truncated large-$n$ part of the canonical partition functions
suffers from a severe sign problem, we confirmed that the zeros relevant to the RW phase transitions 
are insensitive to the truncated terms.

A further question still arises as to how the infinite sum of the truncated terms 
affects the location of Lee-Yang zeros, or whether the fugacity polynomial really converges. 
We \comment{claim} that Lee-Yang zeros near the unit circle are not affected 
even in the limit $n_0 \to \infty$
in which the cutoff is removed. 
To support this claim, we estimate the magnitude of the truncated terms for a simple case
where $Z_n$ is well approximated by a Gaussian function $Z_n^{(\rm Gauss)}$ up to $N$;
namely $\delta_n = Z_n - Z_n^{(\rm Gauss)}\ll 1$ for $|n|< N$, 
and $\delta_n$ is not negligible for $|n|\ge N$. 
Denoting $Z=\sum_n Z_n \xi^n$, and $G=\sum_n Z_n^{(\rm Gauss)} \xi^n$, 
the deviation is given by $Z-G \, \simeq\, \sum_{|n|\ge N} (Z_n - Z_n^{(\rm Gauss)}) \xi^n$. 
For $\xi$ real and negative, $\xi^n$ is positive for even $n$ and negative for odd $n$. 
Our lattice data exhibited in Fig.~\ref{Fig:2014Sep03fig1}
suggests that $\delta_n$ is dominated by $Z_n$, which decreases \comment{monotonically}
as long as there is no phase transition at high temperature.
Then the individual deviation  $\delta_n$ monotonously decreases as $n$ increases, 
so that the overall deviation is bounded as $|Z-G|\le \delta_N |\xi|^N$ due to the cancellation 
between even and odd terms.
The point is that the sum of the truncated terms is bounded by the deviation at $n=N$, 
and does not increase as $N$ is increased.
}
\subsection{Discussion}
\label{sec:discussion}
\comment{Our results} have several implications for theoretical and experimental studies.
The connection between the Gaussianity of the canonical partition functions
and the RW phase transition is worth emphasizing\comment{; }
the former can be extracted from an experimentally
measurable quantity, and the latter is a phenomenon specific to the QGP phase.
\comment{
Here we emphasize that the Gaussianity of the canonical partition function 
is a sufficient condition, but not a necessary one, for the existence of the RW phase transition; if
the canonical partition function is the Gaussian with regard to 
the baryon number, then it implies the existence of the RW phase transition. 
However, the converse is not always true; the Gaussian distribution is 
merely one type of realization of the RW phase transition.}
Although there have been many studies on QCD at the imaginary chemical potential and the RW phase transition~\cite{D'Elia:2014poslat,deForcrand:2002ci,D'Elia:2002gd,D'Elia:2004at,D'Elia:2007ke,D'Elia:2009tm,Cea:2010md,deForcrand:2010he,Nagata:2011yf,Cea:2012ev,Bonati:2014kpa,Kashiwa:2013rm,Sakai:2008py,Morita:2011jva},
\commentb{we are not aware of any literature presenting a way to relate the RW phase transition to 
the canonical partition functions, which can be constructed by measuring number of the baryon in 
the system.}
We predict that the canonical partition functions obtained from the probability
distribution of the net baryon number inferred from the multiplicity of baryons
(or three-quark states) created at high temperature
be well approximated by the Gaussian function and associate the RW phase transition.
\comment{Such measurements, however, may be difficult at present
because observed hadrons in heavy ion collisions are generated at the freeze-out temperature.}
This observation, together with HRG,
might serve as a basis to interpret experimental data obtained in BES
experiments and help to distinguish deviations driven by the critical end point.

The distribution of the Lee-Yang zeros indicates that the RW phase transition
persists at $\mu_I/T=\pi/3$ even in the presence of the real part of
the quark chemical potential.
An analytic form of the canonical partition \comment{of} the Lee-Yang zero distribution obtained in this work
can be used as a reference for future finite density QCD studies.

Recently the STAR Collaboration \cite{Adamczyk:2013dal} reported that the net proton multiplicity closely follows
the Skellam distribution for several collision energies and centralities.
The multiplicity of the net baryon number is also approximately given by a Skellam
distribution in the HRG model~\cite{BraunMunzinger:2011dn}.
In \cite{Nakamura:2013ska}, we applied the Lee-Yang zero theorem to
the net proton multiplicity data in BES experiments, and found that
they did not imply the RW phase transition.
This
is consistent with the common understanding that the freeze-out temperature is lower 
than the temperature at which the RW phase transition takes place.
However, there may be several controversies in deriving the above conclusion, namely, 
the assumption of the equilibrium \comment{or} the use of the net proton multiplicity 
as a substitute for net baryon multiplicity~\cite{Hatta:2003wn,Kitazawa:2012at}.
In addition, the probability distribution has so far been measured for a limited 
range of the net proton number. 

One of the interesting topics is an end point of the RW phase 
transition~\cite{deForcrand:2010he,D'Elia:2009qz,Bonati:2014kpa}. 
The RW-like behavior appears when the free energy approaches
the quartic function at high temperature,
while it does not at $T\approx T_c.$
In this sense, the RW phase transition is likely an indication of the 
completion of the deconfinement transition. 
It may be interesting to examine the relation between the RW end point and 
canonical partition functions, which may provide us with a possibility to study 
the latter experimentally. We leave this problem for future study.

Admittedly, we need to clarify the subtleties involved in numerical
evaluation of canonical partition functions and Lee-Yang zeros. 
The present QCD simulation inevitably contains several lattice artifacts
originated from coarse lattices, large quark masses, and small lattice volumes.
However, we consider that the present results \comment{are} robust and \comment{are} likely to 
hold for another lattice setup in general,
since the RW-like behavior is based only on a few assumptions, 
namely, on the quartic form of the free energy. 


Our numerical results suggest the deviation \comment{of} the Lee-Yang zeros from the RW-like 
behavior at large $\mu$, 
\comment{where the analytic calculation also breaks down.}
We do not understand if this deviation is physical or not
as it lies beyond the applicable range of the present work.
Since one ordinarily expects that there is no phase transition for the quark chemical potential
in the QGP phase, 
we conjecture that the behavior smoothly changes from
the RW-like one to 
a region in which the $c_4$ term in Eq.~(\ref{Eq:2014Apr24eq1}) dominates.
It may be interesting to investigate whether
there is a nontrivial Lee-Yang zero structure at large $\mu$.

\section{Summary}
In summary, we studied the canonical partition functions and Lee-Yang zeros in QCD at high temperature.
We analytically derived them from the free energy in the 
Stefan-Boltzmann limit using the saddle point approximation. 
The canonical partition functions in QCD follow the Gaussian function at high temperature
and at small chemical potential.
We pointed out that the grand canonical partition function is approximately expressed 
as a Jacobi theta function, which enables us to determine all
Lee-Yang zeros analytically. 
These Lee-Yang zeros are located on the negative real axis on the complex plane 
of the baryon fugacity. They are translated into three 
radial lines on the complex plane of the quark fugacity 
owing to the RW periodicity. The zeros exhibit the first-order RW phase transition. 
We also performed lattice QCD simulations. To remove numerical subtleties, we
examined the convergence of the fugacity polynomial and performed the bootstrap analysis of the 
distribution of Lee-Yang zeros. 
The analytic calculations well explain the results obtained from the lattice QCD simulations. 

The novelties of the present study are the analytic solution of the canonical partition functions and  
Lee-Yang zeros, examination of the convergence of the fugacity polynomial, and bootstrap 
analysis for the distribution of Lee-Yang zeros. 
Additionally, we pointed out that the gas of free fermions
provides an exceptional case of the Lee-Yang zero circle theorem. 

We leave some problems for future studies\comment{: } Namely, 
the confirmation of nontrivial 
behavior of Lee-Yang zeros observed at large quark chemical potentials
and the determination of the RW end point in the canonical approach
are worth pursuing.

\begin{acknowledgements}
K. N. thanks Sinya Aoki, Teiji Kunihiro, Akira Ohnishi for discussions, 
Etsuko Itou for advice on error analysis and analytic calculations, 
and Shoji Hashimoto for comments on an early version of the manuscript.
This work is supported in part by JSPS Grants-in-Aid for Scientific Research (Kakenhi) 
Grants No.\ 
00586901 (K. N.),
No. 26-1717 (K. K.),
No. 24340054 and No. 26610072 (A. N.), and
No. 25400259 (S. M. N.).
K. N. is also supported by MEXT SPIRE and JICFuS.
The lattice simulations were mainly performed on SX9 at RCNP and CMC at Osaka University.  
The error analysis was done on the HPC system at RCNP. 
This work is 
also supported by HPCI System Research project (hp130058)
and RICC system at RIKEN.
\end{acknowledgements}

\appendix
\section{Fourier Integral}
\label{Sec:incompletegamma}
A simple way to verify Eq.~(\ref{Eq:2014Apr21eq1}) is to expand the Fourier integral for $Z_n$ as
\begin{align}
Z_n\propto \int_{-\pi/3}^{\pi/3} d\theta\,e^{-a\theta^2}\cos n\theta
=\sum_{k=0}^\infty  \frac{(-1)^k n^{2k}}{(2k)!}I_k~.
\label{A1}
\end{align}
Here $a=V T^3 c_2$, and $I_k$ is defined by
\begin{align}
I_k=\int_{-\pi/3}^{\pi/3} d\theta\,e^{-a\theta^2}\theta^{2k},
\end{align}
which is expressed in terms of complete and incomplete Gamma functions as
\begin{align}
I_k = \frac{1}{a^{k+1/2}}\Bigl(\Gamma(k+1/2)-\Gamma(k+1/2,a \pi^2/9)\Bigr).
\end{align}
Since the incomplete gamma function
$\Gamma(z,p)=\int_p^\infty e^{-t} t^{z-1}dt$
exponentially approaches zero as $p\propto V \to \infty$,
$I_k$ is expressed solely as a complete gamma function.
By using an identity 
\begin{align}
\frac{\Gamma(k+1/2)}{(2k)!}=\frac{\sqrt{\pi}}{4^k k!}~,
\end{align}
Eq.~(\ref{A1}) sums up to the exponential in Eq.~(\ref{Eq:2014Apr21eq1}).


\end{document}

%% file: Texs/shorthand.tex
\newcommand{\beq}{\begin{eqnarray}}   \newcommand{\eeq}{\end{eqnarray}}

  \newcommand{\tr}{\mbox{tr}}
  

%% file: Texs/abb_info
\newcommand{\affSNU}{
Graduate School of Science and Engineering,
Shimane University, Matsue 690-8504, Japan}
\newcommand{\affYITP}{Yukawa Institute for Theoretical Physics, Kyoto University,
Kitashirakawa Oiwakecho, Sakyo-ku, Kyoto 606-8502, Japan}
\newcommand{\affKEK}{KEK Theory Center, High Energy Accelerator Research Organization (KEK), 
Tsukuba 305-0801, Japan}
\newcommand{\affHURIISE}{Research Institute for Information Science and Education, 
Hiroshima University, Higashi-Hiroshima 739-8527, Japan
}